\documentclass[12pt]{article}
\usepackage{amssymb}

\begin{document}

\title{Sufficient conditions for the anti-Zeno effect}

\author{Pavel Exner}

\date{\small Nuclear Physics Institute, Czech Academy of Sciences,
25068 \v{R}e\v{z} near Prague, \\
and Doppler Institute, Czech Technical University, B\v{r}ehov\'{a}
7, 11519 Prague, Czech Republic}

\maketitle

\begin{abstract}
\noindent The ideal anti-Zeno effect means that a perpetual
observation leads to an immediate disappearance of the unstable
system. We present a straightforward way to derive sufficient
conditions under which such a situation occurs expressed in terms
of the decaying states and spectral properties of the Hamiltonian.
They show, in particular, that the gap between Zeno and anti-Zeno
effects is in fact very narrow.
\end{abstract}

\noindent The Zeno effect which means that an unstable system will
never decay if we monitor its decay perpetually is known for
decades. For the first time it was formulated explicitly in this
context by Beskow and Nilsson \cite{BN} and soon after a
mathematical analysis \cite{Ch,Fr} revealed sufficient conditions
under which it exists; it became truly popular after the authors
of \cite{MS} coined its present name. Recently the effect
attracted a new wave of mathematical \cite{EI,EIN,MaS,Sch} and
physical \cite{FMP,FNN,FNP,FPS,HNM,MNP} interest; in the mentioned
papers one can find a more complete bibliography.

Although the opposite situation, in which a frequent measurement
can on the contrary speed up the decay, or ideally to lead to an
immediate disappearance of the unstable system, was also noticed
early \cite{CSM}, it attracted attention only recently -- see,
e.g., \cite{ASW,AEG,KK,RK} and also \cite{Sch} and references
therein. As in the case of the Zeno effect, the problem can be
tackled from two points of view. The more practical one concerns
the increase of the measured lifetime in case when the measurement
are performed with a certain frequency. On the other hand,
theoretically one can ask what happens if the period between two
successive measurements tend to zero. The distinction between the
two is important, because in typical examples, where the spectral
distribution differs from the one giving exponential decay by an
energy cut-off, the decay law unperturbed by the measurements
oscillates around an exponential function. In such a case the mean
value of energy is finite and the system exhibits Zeno effect for
continual observation but at finite measurement frequencies it
switches between Zeno-type and anti-Zeno-type behaviors. We will
thus use the label anti-Zeno only for the infinite-frequency limit
having in mind that validity of such idealizations ``is the heart
and soul of theoretical physics and has the same fundamental
significance as the reproducibility of experimental data'' as
Bratelli and Robinson once put it \cite{BRo}.

Conditions under which the anti-Zeno effect occurs were discussed
in the above mentioned papers. In particular, a necessary and
sufficient condition formulated in a probabilistic language was
derived in \cite{AEG} and reproduced in the review paper
\cite{Sch}. Our aim in the present letter is to present an
alternative simple derivation a sufficient condition given purely
in spectral terms -- see relations (\ref{suffic1}),
(\ref{suffic1'}) below. Moreover, our argument will work also in
the situation when the unstable system has internal degrees of
freedom and more than a simple Fourier transform is needed to
express the decay law -- see (\ref{suffic}) and (\ref{suffic'}).
The same approach give us also a fresh look at the Zeno effect and
shows that the gap between it and the anti-Zeno effect is in fact
very narrow.

We will work within the general framework of quantum kinematics of
a decaying system \cite{Ex}, in other words, we base our
discussion on three objects: a Hilbert space ${\cal H}$
corresponding to an isolated system describing the unstable system
together with its decay products, a projection $P$ specifying a
subspace of ${\cal H}$ referring to the unstable system alone, and
a unitary evolution $e^{-iHt}$ on ${\cal H}$ corresponding to a
self-adjoint total Hamiltonian $H$. We exclude the trivial
situation, of course, assuming that the subspace $P{\cal H}$ is
\emph{not} invariant under $e^{-iHt}$ for $t>0$.

For simplicity we will consider pure states only. If the system is
prepared at the initial instant $t=0$ in a state $\psi\in P{\cal
H}$, its \emph{decay law} unperturbed by measurements, or
non-decay probability at a later time $t$, is
\begin{equation}\label{declaw}
  P(t)= \| P\,e^{-iHt}\psi\|^2,
\end{equation}
in particular, $P(t)= |(\psi, e^{-iHt}\psi)|^2$ if $\dim P=1$
(this quantity has always a time argument so it cannot be confused
with the projection $P$). In the general case the decay law should
be labelled by the initial state, $P_\psi(t)$, but we will avoid
it if there no danger of misunderstanding. If we perform non-decay
measurements at times $t/n,\, 2t/n\,\, \dots,\, t$, all with the
positive outcome, the resulting non-decay probability is
\begin{equation}\label{measured}
  M_n(t)= P_{\psi}(t/n) P_{\psi_1}(t/n) \cdots
  P_{\psi_{n-1}}(t/n)\,,
\end{equation}
where $\psi_{j+1}$ is the normalized projection of $e^{-iHt/n}
\psi_j$ on $P{\cal H}$ and $\psi_0:= \psi$, in particular,
\begin{equation}\label{measured2}
  M_n(t)= (P_{\psi}(t/n))^n
\end{equation}
if $\dim P=1$. Combining the last relation with the fact that
$\lim_{n\to\infty} (f(t/n)^n= \exp\{{-\dot f(0+)t}\}$ whenever
$f(0)=1$ and the one-sided derivative $\dot f(0+)$ exists we see
that the Zeno effect, i.e. $M(t):=\lim_{n\to\infty} M_n(t)=1$ for
all $t>0$, and its anti-Zeno counterpart, i.e. $M(t)=0$ for any
$t>0$, require that $\dot P(0+)$ is zero and negative infinite,
respectively. The same is true if $\dim P>1$ provided the
derivative $\dot P_\psi(0+)$ has such a property for \emph{any}
$\psi\in P{\cal H}$.

It is thus crucial to estimate the quantity $1-P(t)$, or more
explicitly $(\psi, P\psi)- (\psi, e^{iHt}P e^{-iHt}\psi)$, to find
its behavior for small values of $t$. It is easy to see that we
can rewrite it in the form
\begin{equation}\label{diff1}
  1\!-\!P(t)= 2\,\mathrm{Re\,} (\psi, P(I\!-\!e^{-iHt})\psi)
  -\| P(I\!-\!e^{-iHt})\psi\|^2
\end{equation}
The left-hand side of (\ref{diff1}) can be expressed as
\begin{equation}\label{diff2}
  4\int_{-\infty}^\infty \!\sin^2 \frac{\lambda t}{2}\,
  d\|E^H_\lambda \psi\|^2 -4 \left\|\int_{-\infty}^\infty
  \!e^{-i\lambda t/2}\,\sin \frac{\lambda t}{2}
  \, dP E^H_\lambda \psi\, \right\|^2
\end{equation}
if we use the spectral representation of $e^{-iHt}$ in terms of
the spectral measure $E_H$ of the Hamiltonian $H$ (generated by a
non-decreasing projection-valued function $\lambda \mapsto
E^H_\lambda: = E_H((-\infty,\lambda])$). By Schwarz inequality the
quantity (\ref{diff2}) is non-negative; our aim is to find tighter
upper and lower bounds for it.

Let us begin with the case $\dim P=1$ when (\ref{diff2}) becomes
\begin{equation}\label{diff3}
  4\int_{-\infty}^\infty \!\sin^2 \frac{\lambda t}{2}\,
  d\omega(\lambda) -4 \left|\int_{-\infty}^\infty
  \!e^{-i\lambda t/2}\, \sin \frac{\lambda t}{2}
  \, d\omega(\lambda) \right|^2
\end{equation}
with $d\omega(\lambda):= d(\psi, E^H_\lambda \psi)$. In most decay
models $\psi$ belongs to the absolutely continuous spectral
subspace of $H$ in which case one has $d\omega(\lambda)=
\omega(\lambda) d\lambda$ with some $\omega\in L^1$, however, we
will not need this assumption. Using the spectral-measure
normalization, $\int_{-\infty}^\infty d\omega(\lambda)=1$, we can
rewrite the difference (\ref{diff3}) as
\begin{eqnarray*}
  \lefteqn{2\int_{-\infty}^\infty \int_{-\infty}^\infty \left(
  \sin^2 \frac{\lambda t}{2}+ \sin^2 \frac{\mu t}{2} \right)
  d\omega(\lambda) d\omega(\mu)} \\ &&
  - 4\, \int_{-\infty}^\infty \int_{-\infty}^\infty
  \cos \frac{(\lambda-\mu) t}{2} \,
  \sin \frac{\lambda t}{2}\, \sin \frac{\mu t}{2}\,
  d\omega(\lambda) d\omega(\mu)
\end{eqnarray*}
or
\begin{equation}\label{diff4}
  1\!-\!P(t)= 2 \int_{-\infty}^\infty \int_{-\infty}^\infty
  \sin^2 \frac{(\lambda-\mu) t}{2}\,
  d\omega(\lambda) d\omega(\mu)
\end{equation}
Take $\alpha\in (0,2]$. Using the inequalities $|x|^\alpha \ge
|\sin x|^\alpha \ge \sin^2 x$ together with $|\lambda-\mu|^\alpha
\le 2^\alpha (|\lambda|^\alpha +|\mu|^\alpha)$ we infer from the
expression (\ref{diff4}) that
\begin{eqnarray*}
  \lefteqn{ \frac{1-P(t)}{t^\alpha} \le 2^{1-\alpha}
  \int_{-\infty}^\infty \int_{-\infty}^\infty
  |\lambda-\mu|^\alpha d\omega(\lambda) d\omega(\mu) } \\ &&
  \phantom{AA}
  \le 2 \int_{-\infty}^\infty \int_{-\infty}^\infty
  (|\lambda|^\alpha +|\mu|^\alpha) d\omega(\lambda) d\omega(\mu)
  \\ && \phantom{AA} \le 4 \langle |H|^\alpha \rangle_\psi.
\end{eqnarray*}
This means that $1-P(t)= {\cal O}(t^\alpha)$ if $\psi\in {\rm
Dom\,} (|H|^\alpha)$. If this is true for some $\alpha>1$ we have
naturally Zeno effect, although this requirement is slightly
stronger than the other known sufficient conditions; recall that
$\dot P_\psi(0+)=0$ holds whenever $\langle |H| \rangle_\psi$ is
finite as it is known for longtime, see \cite{HE},
\cite[Thm.~1.3.1]{Ex}, and also \cite{Fr}.

On the other hand, by negation we infer that $\psi\not \in {\rm
Dom\,} (|H|)$ is a \emph{necessary condition} for the
(one-dimensional) anti-Zeno effect. Notice that in case of the
absolutely continuous spectrum this necessary condition follows
also from the Lipschitz regularity, since $P(t)=
|\hat\omega(t)|^2$ and $\hat\omega$ is bounded and uniformly
$\alpha$-Lipschitz \emph{iff} $\int_\mathbb{R} \omega(\lambda) (1+
|\lambda|^\alpha)\,d\lambda < \infty$.

Some may believe that by this the problem is closed -- it is a
common mistake that only states from the domain of the Hamiltonian
make physical sense. In reality one can never test experimentally
that a given does \emph{not} belong to $D(H)$, see
\cite[Sec.~I.6]{Ex}. To find a \emph{sufficient condition} let us
observe that for $\lambda, \mu\in [-1/t,1/t]$ there is a positive
$C$ independent of $t$ such that
\begin{equation}\label{lowest}
  \left| \sin \frac{(\lambda-\mu) t}{2} \right| \ge
  C|\lambda-\mu|t\,;
\end{equation}
one can make the constant explicit but it is not necessary.
Consequently, we have the estimate
$$ 
  1-P(t) \ge 2C^2t^2 \int_{-1/t}^{1/t} d\omega(\lambda)
  \int_{-1/t}^{1/t} d\omega(\mu) (\lambda-\mu)^2
$$ 
which in turn implies
\begin{equation}\label{difflow}
  \frac{1-P(t)}{t} \ge 4C^2t \left\{
  \int_{-1/t}^{1/t} \lambda^2\, d\omega(\lambda)
  \int_{-1/t}^{1/t} d\omega(\lambda)
  -\left( \int_{-1/t}^{1/t} \lambda\, d\omega(\lambda)
  \right)^2 \right\}.
\end{equation}
The anti-Zeno effect occurs if the the right-hand side diverges as
$t\to 0$ which is true if
\begin{equation}\label{suffic1}
  \int_{-N}^N \lambda^2\, d\omega(\lambda)
  \int_{-N}^N d\omega(\lambda)
  -\left( \int_{-N}^N \lambda\, d\omega(\lambda)
  \right)^2 \ge c N^\alpha
\end{equation}
holds for any $N$ and some $c>0,\, \alpha>1$, or slightly more
generally, if the inverse of the right-hand-side expression in
(\ref{suffic1}) behaves like $o(N)$ as $N\to\infty$.

The obtained sufficient condition can be also written in a
slightly more compact form if we introduce the operators
$H_N^\beta:= H^\beta E_H(\Delta_N)$ with the spectral cut-off to
the interval $\Delta_N:=(-N,N)$, in particular, we denote $I_N:=
E_H(-N,N)$. In this notation, the inequality (\ref{suffic1})
becomes
\begin{equation}\label{suffic1'}
  \langle H^2_N\rangle_\psi \langle I_N\rangle_\psi
  - \langle H_N\rangle^2_\psi \ge c N^\alpha\,.
\end{equation}
Let us stress that the condition which we have derived does
\emph{not} require the Hamiltonian $H$ to be unbounded from below
as it is the case with the exponential decay law; to satisfy
(\ref{suffic1}) it is enough that the spectral distribution has a
slow decay in one direction only.

As an \emph{example}, consider a Hamiltonian bounded from below
and $\psi$ from its absolutely continuous spectral subspace such
that the corresponding distribution function behaves as
$\omega(\lambda) \approx c\lambda^{-\beta}$ as $\lambda\to
+\infty$ for some $c>0$ and $\beta\in(1,2)$. While $\int_{-N}^N
\omega(\lambda)\, d\lambda$ tends to one as $\lambda\to +\infty$,
the other two integrals diverge giving
$$ 
  cN^{2-\beta}-c^2N^{4-2\beta}
$$ 
as the asymptotic behavior of the left-hand side, where the first
term is dominating; it gives $\dot P(0+)=-\infty$ so the anti-Zeno
effect occurs. The above argument shows that in the same situation
$\beta>2$ leads to Zeno effect; this shows that the exponential
decay law indeed walks a thin rope between the Scylla of
{\ae}ternal preservation and Charybda of immediate destruction. Of
course, the exponential decay appears only if the spectrum of $H$
is the whole real line. For a \emph{semibounded} $H$ with the
asymptotic behavior $\omega(\lambda) \approx c\lambda^{-1}$ the
reduced evolution $(\psi, e^{-iHt}\psi)$ typically exhibits rapid
oscillations around $t=0$ which may obscure existence of the Zeno
limit -- cf.~\cite{Ex}, Rem.~2.4.9.

Let us show next how the situation looks like when the unstable
system is allowed to have internal degrees of freedom, $\dim P>1$.
One might expect that the sufficient condition is given by
(\ref{suffic1'}) again but this guess is wrong. To find the
answer, we denote by $\{\chi_j\}$ an orthonormal basis in $P{\cal
H}$; it allows to write the second term in (\ref{diff2}) as
$$ 
  -4 \sum_m \left|\int_{-\infty}^\infty
  \!e^{-i\lambda t/2}\, \sin \frac{\lambda t}{2}
  \, d(\chi_m, E^H_\lambda \psi) \right|^2.
$$ 
We also expand the initial state vector $\psi$ as $\psi= \sum_j
c_j\chi_j$ with $\sum_j |c_j|^2 =1$ and denote
$d\omega_{jk}(\lambda):= d(\chi_j, E^H_\lambda \chi_k)$, which is
naturally a real-valued measure symmetric with respect to
interchange of the indices. Since the other measure appearing in
(\ref{diff2}) can be written as
$$ 
  d\|E^H_\lambda\psi\|^2 = \sum_{jk} \bar c_j c_k
  d\omega_{jk}(\lambda)\,,
$$ 
we can cast the quantity of interest into the form
\begin{eqnarray}\label{diff5}
  \lefteqn{1-P(t)= 4\sum_{jk} \bar c_j c_k \bigg\{
  \int_{-\infty}^\infty \!\sin^2 \frac{\lambda t}{2}\,
  d\omega_{jk}(\lambda) } \nonumber \\ &&
  - \sum_m \int_{-\infty}^\infty
  \! e^{-i\lambda t/2}\,\sin \frac{\lambda t}{2}
  \, d\omega_{jm}(\lambda)\, \int_{-\infty}^\infty
  \! e^{i\mu t/2}\,\sin \frac{\mu t}{2} \, d\omega_{km}(\mu) \bigg\}.
\end{eqnarray}
If $\dim P=\infty$ one has to check, of course, the convergence of
the series used in the argument and correctness of interchanging
of the summation and integration which is easily done by means of
Parseval relation.

In the next step we employ normalization of the spectral measure
which gives $\int_{-\infty}^\infty d\omega_{jk}(\lambda) =
\delta_{jk}$. It is then a straightforward exercise to rewrite the
curly bracket and to show that the right-hand side of
(\ref{diff5}) can be rewritten as
\begin{equation} \label{diff6}
1-P(t)= 2\sum_{jkm} \bar c_j c_k
  \int_{-\infty}^\infty \int_{-\infty}^\infty
  \sin^2 \frac{(\lambda-\mu) t}{2}\,
  d\omega_{jm}(\lambda) d\omega_{km}(\mu)\,.
\end{equation}
Returning to the projection-valued measures we can write the
right-hand side of (\ref{diff6}) also concisely as
$$ 
  2 \int_{-\infty}^\infty \int_{-\infty}^\infty
  \sin^2 \frac{(\lambda-\mu) t}{2}\,
  (\psi, dE^H_\lambda P dE^H_\mu \psi)\,.
$$ 
To get a lower bound to the left-hand side of (\ref{diff6}) we
employ again the inequality (\ref{lowest}) which gives
\begin{eqnarray*}
  \lefteqn{1-P(t)\ge 2C^2t^2
  \int_{-1/t}^{1/t} \int_{-1/t}^{1/t}
  (\lambda-\mu)^2\,
  (\psi, dE^H_\lambda P dE^H_\mu \psi)}
  \\ && =4C^2t^2 \bigg\{
  \int_{-1/t}^{1/t} \int_{-1/t}^{1/t}
  \lambda^2\, (\psi, dE^H_\lambda P dE^H_\mu \psi)
  \\ && \phantom{AA}
  - \int_{-1/t}^{1/t} \int_{-1/t}^{1/t}
  \lambda\mu\, (\psi, dE^H_\lambda P dE^H_\mu \psi)
  \bigg\} \\ &&
  =4C^2t^2 \left\{ (\psi, H^2_{1/t} P I_{1/t} \psi)
  - \|PH_{1/t}\psi\|^2 \right\}\,, \phantom{AAAAAAA}
\end{eqnarray*}
where the symbol $H_b$ denotes again the cut-off Hamiltonian,
$HE_H(\Delta_b)$ with $\Delta_b:= (-b,b)$. Hence a sufficient
condition for the anti-Zeno effect to occur in this more general
situation is, for instance,
\begin{equation}\label{suffic}
  \langle H^2_N P I_N\rangle_\psi
  - \|PH_N\psi\|^2 \ge c N^\alpha
\end{equation}
for some $c>0$ and $\alpha>1$, both independent of $\psi$, as a
proper generalization of the one-dimensional condition
(\ref{suffic1'}) -- notice that the second term at the left-hand
side of (\ref{suffic}) can be also written as $\langle
H_NPH_N\rangle_\psi$ -- or slightly more generally
\begin{equation}\label{suffic'}
  \left(\langle H^2_N P I_N\rangle_\psi
  - \|PH_N\psi\|^2\right)^{-1} = o(N)
\end{equation}
as $N\to\infty$ uniformly w.r.t. $\psi\in P{\cal H}$. The meaning
of these conditions is similar as before: the energy distribution,
now for any possible state of the unstable system, must be
sufficiently spread to ensure that the initial decay rate of such
a $\psi$ is infinite.

Let us add some comments. First we observe that in the case $\dim
P>1$ a system subject to a perpetual observation can exhibit also
a more complicated behaviour. A simple example is a combination of
Zeno and anti-Zeno effect. Take $\{{\cal H}_j,P_j,H_j\}, j=1,2,$
with $\dim P_j=1$ such that $\dot P_1(0+)=0$ and $\dot
P_2(0+)=-\infty$, and consider the combined system described by
the triple $\{{\cal H}_1 \oplus{\cal H}_2,P_1\oplus P_2,H_1\oplus
H_2\}$. If the initial state of such an unstable system is
represented by a non-trivial linear combination $\psi= c_1\psi_1+
c_2\psi_2$ and we monitor it continuously, the second component of
$\psi$ disappears immediately while the first one will survive
forever. Of course, one can imagine various more complicated
combinations, especially in the case when $\dim P$ is infinite.

Another comment concerns physical relevance of the conditions
discussed here. The point is that they involve asymptotic behavior
of spectral distributions at high energies, i.e. something which
cannot be in principle verified experimentally. A similar question
arises also in connection with the Zeno effect, of course, but
there at least sometimes we can be sure that the needed moment of
the spectral distribution exists, in particular, if the experiment
involves a filtering into a finite energy window, while checking
the divergence of such integrals is \emph{always} out of
experimental reach as we have remarked above.

What one can verify, however, is whether an energy distribution of
a state coincides with with the one leading to the anti-Zeno
effect, for instance, decreasing as $c\lambda^{-\beta}$ with some
$\beta\in (1,2)$ over a wide range of energies up to some value
$\lambda_c$. If this is the case the difference of the theoretical
and actual $P(t)$ will be small and the difference in the initial
behavior will be significant at the time scale characterized by
$\hbar\lambda_c^{-1}$, hence with the measurement frequencies
small enough at this scale one should be able to observe a
significant reduction of the measured lifetime, demonstrating the
anti-Zeno effect practically.

In conclusion, we have derived sufficient conditions under which
unstable systems exhibit anti-Zeno effect using nothing else than
properties of spectral distribution of the decaying states. The
conditions impose restrictions neither on the lower bound of the
spectrum of the corresponding Hamiltonian nor on the dimension of
the unstable system subspace.

\subsection*{Acknowledgments}

The author is grateful to the referee who spotted an error in an
earlier version of the paper. The work was supported by Czech
Academy of Sciences and its Grant Agency within the projects
A100480501 and IRP AV0Z10480505.

\end{document}